\RequirePackage[2020-02-02]{latexrelease}
\documentclass[preprint,showpacs,showkeys,preprintnumbers]{revtex4}

\usepackage{graphicx}
\usepackage{dcolumn}
\usepackage{bm}
\usepackage{epsfig}
\usepackage{epstopdf}
\usepackage{grffile}
\usepackage{color}
\usepackage{colordvi}
\usepackage{amssymb}
\usepackage{rotating}
\usepackage{lscape}
\usepackage{float}
\usepackage{mathtools}
\usepackage{footnote}
\makesavenoteenv{environmentname}
\usepackage{hyperref}

\usepackage[T1]{fontenc} 

\begin{document}
\title{Constraining New Physics by Studying Single Top Quark Production }

\author{Maryam Roknabadi}
\email{m.roknabady@stu.yazd.ac.ir}
\affiliation{Department of Physics, Yazd University, P.O. Box 89195-741, Yazd, Iran}

\author{Saeid Paktinat Mehdiabadi}
\email{spaktinat@yazd.ac.ir}
\affiliation{Department of Physics, Yazd University, P.O. Box 89195-741, Yazd, Iran}

\author{Seyed Mohammad Reza Tabatabaei}
\email{mr.tabatabaei@outlook.com}
\affiliation{Independent Researcher}
	
\date{\today}

\begin{abstract}
This study introduces a beyond the Standard Model framework by adding a charged scalar particle, \( H_{c} \), analogous to the charged Higgs in the Two-Higgs-Doublet Model (2HDM), but analyzed without the restrictive constraints typically applied in 2HDM scenarios.
To assess the impact of  \( H_{c} \) on single top quark production, two approaches are taken. First, variations in the production cross-section via the $s$ and $t$-channels are examined under different values of the model parameters, compared to the SM predictions. Second, spin correlations are analyzed, focusing on differences in the spin of the mediated particle between the new model and SM in the single top quark production via the $s$-channel.
Within the framework of the simplified phenomenological model considered here, a significant fraction of the parameter space can be constrained by combining the two approaches.
\end{abstract}
\maketitle

\section{Introduction}\label{sec.intr}
The fundamental particles recognized in nature are described as quantum fields governed by interactions among them. The Standard Model  of the elementary particles (SM) is a quantum gauge field theory that successfully explains these interactions \cite{ParticleDataGroup:2024cfk}. Among the fermions, the last quark discovered within the SM framework is the top quark, which, due to its large mass (173 GeV) and extremely short lifetime ($5\times10^{-25}$s), serves as a unique probe for testing SM and exploring physics beyond it \cite{ParticleDataGroup:2024cfk}.
In other words, precise measurements of the top quark's properties can reveal potential deviations from the predictions of SM. Furthermore, if a fourth generation of quarks exists, these precise measurements could provide indications of it, playing a fundamental role in exploring and studying new physics.
The top quark decays before hadronization, allowing its spin and other quantum properties to be directly transferred to the decay products. This feature makes it an ideal candidate for spin correlation studies, which can reveal subtle effects such as CP violation or anomalous couplings at the $Wtb$ vertex \cite{CMS:2024vqh}, \cite{schulte2017top}. Recent measurements at LHC \cite{Evans:2008zzb} have focused on polarization asymmetries and $W$ boson helicity fractions, using effective field theory (EFT) frameworks to interpret deviations from SM predictions \cite{schulte2017top}. During the decay of the top quark, a $W$ boson is produced. This boson can appear in three helicity configurations, left-handed, right-handed, or longitudinal (i.e., spinless along its momentum direction). The proportion corresponding to each configuration is referred to as the helicity fractions of the $W$ boson. Measuring these fractions helps to determine whether the behavior of the $W$ boson aligns with the SM predictions.
On the other hand, polarization asymmetries refer to differences in the helicity orientations of the decay products. These asymmetries may serve as indirect indicators of the presence of anomalous couplings or potential CP violation, offering insight into the physics beyond SM.
Top quarks are produced predominantly via top-antitop pair production through strong interactions, but also singly via electroweak processes in the $t$-channel, $s$-channel, and $tW$-associated production. Among these, the $t$-channel has the highest cross-section and is sensitive to the $V_{tb}$ element of the CKM matrix \cite{CMS:2024vqh}, \cite{andrea2023single}. The s-channel, though rarer, provides a clean environment to probe new resonances such as $W'$ bosons or effects from extra dimensions \cite{andrea2023single}.
Precise measurements of differential cross-sections in these channels have enabled constraints on anomalous couplings, flavor-changing neutral currents (FCNCs), and top quark polarization \cite{CMS:2024vqh}, \cite{andrea2023single}. Spin correlation studies in top pair production have also revealed entanglement effects. The recent CMS results showing statistically significant quantum entanglement in high-mass top quark systems \cite{CMS:2024vqh}, \cite{schulte2017top}.
Furthermore, the top quark's mass being close to the electroweak symmetry breaking scale (246 GeV) suggests it may play a role in stabilizing the Higgs potential \cite{ParticleDataGroup:2024cfk}. If a fourth generation of quarks exists, deviations in top quark observables could provide indirect evidence \cite{grannis202125}. Thus, the top quark remains central to both validating the SM and probing new physics.

This work introduces a minimal extension to the Standard Model through a new charged scalar particle, $H_{c}$. Unlike the charged Higgs in conventional 2HDM scenarios, this particle is investigated independently of the full 2HDM constraints. We focus on two complementary probes in single top quark production: precision cross-section measurements and spin correlation analyses.
Previous studies of charged scalar particles at the LHC have primarily focused on direct production modes, such as $ pp  \rightarrow  tbH^{\pm}$ with subsequent decays $ H^{\pm} \rightarrow \tau \nu$, $tb$, or $cs$ \cite{aad2025search}, 
 as well as associated production in top-quark decays $ t \rightarrow  bH^{\pm}$ \cite{kling2015light}. 
In contrast, the present work investigates the indirect effects of a charged scalar singlet appearing as a virtual mediator in single top quark production via the s- and t-channels. Moreover, while most phenomenological analyses rely solely on cross-section measurements, we combine production-rate constraints with a spin-correlation analysis that exploits the different spin nature of the scalar mediator relative to the $W$ boson. This dual approach provides complementary constraints and represents a novel strategy for probing simplified charged-scalar extensions of  SM in the single-top sector \cite{arco2025top}. 
We emphasize that our analysis is performed within a simplified phenomenological framework, and the constraints obtained should be interpreted as indicative of the potential reach of such probes rather than as definitive exclusion bounds for a complete UV-complete theory.

The paper is organized as follows, after this introduction, the model is explained in Sec. \ref{sec.model}. Section \ref{sec.ana} summarizes the approaches used in this study and their results. The paper is concluded in Sec. \ref{sec.con}.

\section{Phenomenological Model}\label{sec.model}
Extending the higgs sector of SM is one of the most popular extensions of SM. Adding a new higgs doublet, known as Two-Higgs-Doublet Models (2HDM) \cite{Branco_2012}, is a famous example. Nowadays, different parts of this model are constrained strongly by the experimental and theoretical conditions. In this analysis, a new charged scalar particle is added to SM. It is somehow similar to the well known charged higgs boson of 2HDM. To relax the different strong constraints, adding a new doublet and being in 2HDM framework is ignored and only adding a new charged scalar particle (\( H_{c}\)), obeying the following Lagrangian  is considered. 
\begin{equation}
L_{H_{c}} = \alpha H_{c} (V_{ud} m_{u} \overline{u} P_{L}d + V_{ud} m_{d} \overline{u}P_{R}d + m_{l}\overline{\nu}_{L}l_{R} ) + H.c.
\label{eq:hcL}
\end{equation}
Where  \(m_{u}\), \(m_{d}\)  and  \( m_{l}\)  are the masses of the u and d quarks and the lepton, respectively. The characters, $u$ and $d$, correspond to the spinors of the $up$-type and $down$-type quarks, respectively. The right-handed and left-handed projection operators are shown as \(P_{R}\) and \(P_{L}\), respectively. The matrix \(V_{ud}\) is the coupling coefficient of the two quarks in the CKM  matrix, and \(\alpha\) is the coupling constant in the considered model.
The particle \(H_{c}\) is introduced as a singly charged scalar singlet under the SM gauge group $ SU(3)_{C} \times SU(2)_{L} \times U(1)_{Y}$, with weak isospin \(T=0\) and hypercharge \(Y=1\). Its kinetic Lagrangian is given by
\begin{equation}
L_{kin} =(D_{\mu} H^+_{c}) (D^{\mu} H^-_{c})
\label{eq:kinL}
\end{equation}
where \(D_{\mu}=\partial_{\mu}+g^{\prime} YB_{\mu} \) is the covariant derivative. Being an \(SU(2)_{L}\) singlet, \(H_{c}\) does not couple directly to $W$ bosons, but couples to the $Z$ boson and the photon through its hypercharge. Such a field content appears, for instance, in the Zee model for radiative neutrino masses \cite{zee1980theory}, in Zee-type models with arbitrary numbers of scalar doublets and singlets \cite{florentino2021off}, and in lepton-portal dark matter extensions \cite{Jueid:2023pjp}. 
The Yukawa interactions in Eq.\ref{eq:hcL} are understood as an effective description arising from the mixing of the singlet \( H_{c}\) with the charged components of the Higgs doublets in the underlying  UV  completion, which yields a coupling structure analogous to that of the Type-I 2HDM charged Higgs. 
The scalar potential of  \( H_{c}\) includes the renormalizable terms 
\begin{equation}
V \supset (m^2_{H_{c}}H^+_{c} H^-_{c}) + \lambda_c(H^+_{c} H^-_{c})^2.
\label{eq:V}
\end{equation}
where $\lambda_c > 0$ ensures that the potential is bounded from below. Since all operators in the Lagrangian of Eq.~\ref{eq:hcL} and in the scalar potential are of dimension four, the model is renormalizable. Perturbative unitarity of the scalar sector is maintained provided the quartic couplings satisfy the constraints derived in multi-scalar models~\cite{florentino2021off}.
These constraints require that the eigenvalues $\Lambda_i$ of the scalar-scalar scattering matrix, evaluated in the s-wave ($\ell=0$) partial wave, satisfy $|\Lambda_i| < 8\pi$.
In the current phenomenological scan, it is ensured that the couplings remain in the perturbative regime ($|\alpha|<\sqrt{4\pi}$, $|\lambda_c|<8\pi$). In Eq.~\ref{eq:hcL}, $u$ and $d$ denote generic up-type and down-type quarks; for the third-generation quarks relevant to single top quark production, the coupling is obtained by the replacement
$V_{ud} \rightarrow V_{tb}$, $m_{u} \rightarrow m_{t}$, and $m_{d} \rightarrow m_{b}$.
It should be emphasized that the present study adopts a simplified phenomenological approach. The Lagrangian in Eq.~\ref{eq:hcL} should be understood as an effective description of a more complete UV completion, such as the Zee model~\cite{zee1980theory}, Zee-type models~\cite{florentino2021off}, or lepton-portal extensions~\cite{Jueid:2023pjp}, where the charged scalar singlet arises naturally. In such completions, additional fields (e.g., a second Higgs doublet or right-handed neutrinos) would be present, and the full scalar potential would include terms that ensure vacuum stability and perturbative unitarity. The current analysis focuses on the collider phenomenology of ($H_{c}$) in the single top quark production, leaving a detailed study of the UV completion and its implications for neutrino masses or dark matter to future work.

Finally, it is noted that the charged scalar singlet, $H_{c}$, does not introduce gauge anomalies, since it is a complex scalar and carries no fermionic quantum numbers. Its interactions respect the SM gauge symmetries at the renormalizable level.

The newly introduced Lagrangian is very similar to the Lagrangian of the charged higgs in the type-I of Two-Higgs-Doublet Models (2HDM)~\cite{barger1990new}.
\begin{equation}
L_{H^{\pm}} = -H^{+} \left( \frac{\sqrt{2}V_{ud}}{\nu}\overline{u}(m_{u}XP_{L} + m_{d}YP_{R})d + \frac{\sqrt{2} m_{l}}{\nu}Z\overline{v}_{L}l_{R}\right) + H.c.,
\label{eq:2hdmL}
\end{equation}
where, $\nu$ represents the vacuum expectation value (VEV), and the values of X, Y, and Z differ across various types of Two-Higgs-Doublet Models (2HDM). For type-I of 2HDM, these values are equivalent to
\begin{equation}
X=Y=Z=\cot\beta.
\label{eq:XYZ}
\end{equation}
The other parameters in the equation are consistent with the parameters in the proposed Lagrangian of the particle $H_{c}$.
By comparing the two Lagrangian equations, the following relation for the coefficient $\alpha$ can be found:
\begin{equation}
\alpha=\frac{\sqrt{2}}{\nu} \cot(\beta).
\label{eq:alpha}
\end{equation}

It should be noted that, in the phenomenological framework of this work, $\alpha$ is treated as an independent effective parameter that characterizes the strength of the Yukawa interaction in Eq.~\ref{eq:hcL}. The relation in Eq.~\ref{eq:alpha}, is not a fundamental prediction of the model; rather, it is introduced solely for numerical implementation purposes. Specifically, the 2HDMC software~\cite{eriksson20102hdmc} 
requires the charged Higgs couplings to be specified in terms of the standard 2HDM parameters, including $\tan\beta$. Equation~\ref{eq:alpha} therefore serves as a computational bridge that maps our independent parameter $\alpha$ onto the $\tan\beta$ input of the software.

In the underlying UV completion (e.g., a Zee-type model with two Higgs doublets and one charged singlet~\cite{florentino2021off}), the physical Yukawa coupling of the singlet-like charged scalar arises from the mixing between the singlet and the charged components of the Higgs doublets. In our simplified phenomenological description, this combination is absorbed into the single effective parameter $\alpha$.


\subsection{Constraints and consistency checks}\label{subsec.II.B.model}
Since the proposed model considers the interaction terms structurally similar to those of the charged Higgs in the Type-I 2HDM, it is necessary to discuss the theoretical and experimental constraints that apply to the parameter space. In what follows, each class of constraints relevant to the introduced simplified phenomenological framework is addressed.
\begin{itemize}
\item Theoretical constraints: The scalar potential of $H_{c}$ includes the renormalizable quartic term $\lambda_c$. To ensure that the potential is bounded from below, $\lambda_c$ is required to be positive~\cite{gonderinger2010vacuum}. Perturbative unitarity of the scalar sector is maintained provided the quartic couplings satisfy $|\Lambda_i|<8\pi$ for all eigenvalues of the scalar scattering matrix~\cite{florentino2021off}. In the phenomenological scan, we restrict the effective coupling to $|\alpha|<\sqrt{4\pi}$ and assume $\lambda_c$ remains perturbative, which guarantees that the theory does not violate perturbative unitarity.
\item Flavour constraints: The Yukawa structure in Eq.~\ref{eq:hcL} is analogous to the Type-I 2HDM. In the Type-I 2HDM, the stringent lower bound $m_{H^{\pm}}\geq 580$ GeV from $BR(B\rightarrow X_{s}\gamma)$ that applies to the Type-II 2HDM is significantly relaxed~\cite{florentino2021off,misiak2017weak}. Other flavour observables, such as $B_{s,d}\rightarrow \mu^+\mu^-$, constrain only the region corresponding to large $\alpha$ (which would map to low $\tan\beta$ in the Type-I 2HDM interface)~\cite{kling2023light}, 
which is already partially excluded by our cross-section analysis.
\item LEP and direct LHC searches: LEP searches for charged scalars set a lower limit of approximately 80 GeV on the mass of a charged Higgs boson in the 2HDM~\cite{buskulic1995search}. 
For a singly charged scalar singlet, however, the LEP bound is relaxed to $m_{H_{c}} \geq 65$ GeV because the pair-production cross section via $\gamma/Z$ and the decay branching ratios differ from those of a doublet~\cite{cao2018searching}. 
Direct LHC searches for charged Higgs bosons primarily target the 2HDM decay channels $(H^{\pm} \rightarrow \tau\nu$, tb, cs) and associated production with top quarks~\cite{aad2025search}. 
These searches do not directly constrain our model, where $H_c$ appears as a virtual mediator in single top production. Nevertheless, the parameter regions excluded by our cross-section and spin-correlation analyses lie well within the mass range probed by LHC searches.
\end{itemize}
In summary, the parameter space explored in this work ($m_{H_c} \in [50, 450]$ GeV, $\alpha \in [0, 0.6]$) is consistent with the theoretical and experimental constraints discussed above within the simplified phenomenological framework adopted in this analysis. A more comprehensive analysis including the full set of constraints from a specific UV completion (e.g., the Zee model) is beyond the scope of this phenomenological study and is deferred to future work.

Provided the model is in place, in the next section, different approaches are used to constrain the new model.

\section{Analysis}\label{sec.ana}
\begin{figure}[!htb]
\centering
\includegraphics[width=0.55\textwidth]{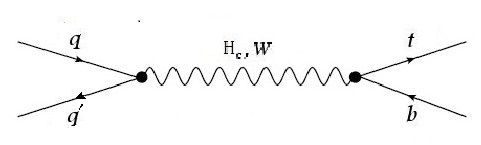}
\includegraphics[width=0.35\textwidth]{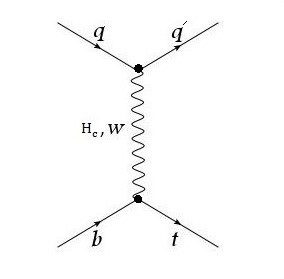}
\caption{ (left) Feynman diagram of single top quark production in the $s$-channel (\(q\bar{q^{\prime}} \to \bar{b}t \)).  (right) Feynman diagram of single top quark production in the $t$-channel (\(qb \to tq^{\prime} \)). The time evolution is from left to right.}
\label{fig:SingleTop}
\end{figure}

To constrain the new model, or find a signature of a new physics beyond SM,  the single top quark production at LHC is studied. In the LHC experiments, protons collide at the center-of-mass energy of 13 TeV. There are different processes producing single top quark in proton-proton (pp) collisions. Here, two main processes known as $s$-channel (\(q\bar{q^\prime} \to \bar{b}t \)) and $t$-channel (\(qb \to  tq^{\prime} \)) are considered. With the inclusion of \( H_{c}\), the production of single top quarks and their antiparticles in the $s$- and $t$-channel occurs through the processes shown in Fig.\ref{fig:SingleTop}.
As expected from the Lagrangian structure, \( H_{c} \) can replace the \( W \) boson and influence the single top quark production. In the following, two different approaches are studied to find the effects of this influence, cross section measurement and spin effects.
To study the effects, the MadGraph v2-6-3 software \cite{alwall2011madgraph} is used to simulate the pp collisions. The cross-sections of single top quark production in Type-I 2HDM via the $s$- and $t$-channels are calculated. The calculations are performed for different  values of the charged higgs mass and $\tan(\beta)$, defining different ($m_{H_c}$) and $\alpha$, respectively. 
To incorporate the new model into this software, the 2HDMC-1-8-0  software \cite{eriksson20102hdmc} is utilized. This software computes the couplings and decay widths of the particles of 2HDM by taking the model parameters as input, enabling their use in MadGraph software.

\subsection{Cross-section constraints}
In the first approach, the effect of adding a new charged scalar to SM is investigated on single top quark production cross-section. Depending on the charged scalar mass ($m_{H_c}$) and its coupling ($\alpha$), single top quark production cross-section can receive different modifications from the proposed new model. Experimental measurements of single top quark production, which occur via electroweak interactions, have not yet reached the required precision. Improving the accuracy of these measurements imposes stronger constraints on physics beyond SM. In this study, the measurement of the single top quark production cross-section is utilized to constrain the new model. Based on the measured values of the single top quark production cross-section, the allowed parameter space of the introduced model is presented. 
To compute the single top quark production cross-section in the \( s \)- and \( t \)-channels, the MadGraph software is utilized, generating 20,000 tree-level events at a center-of-mass energy of \( \sqrt{s} =13 \) TeV.

The impact of all various 2HDM parameters on the cross-sections are examined, and only the mass of the charged scalar particle \( m_{H_c} \) and \( \tan\beta \), are found to significantly affect the cross-section values. Consequently, only these two parameters are varied in the subsequent analysis, while all other 2HDM parameters are kept fixed equal to their default values. The MadGraph software parameters not explicitly mentioned here also retain their default values.
To constrain the calculated cross-sections, two methods have been employed to define the allowed ranges for the single top quark production cross-section. 
The first allowed range is determined based on the latest and most precise experimental measurement of the single top quark production cross-section in the $s$-channel. The ATLAS experiment at CERN reported this measurement in 2023 as $\sigma_{\text{s-channel}} = 8.2^{+3.5}_{-2.9}$ pb \cite{ATLAS:2022wfk}.

The cross-section values calculated by the software are scaled using the ratio,  $K$-factor, of the most precise SM predictions for cross-sections (10.32 pb\cite{ATLAS:2022wfk}) to tree-level cross-sections in SM (6.92 pb) to account for higher-order corrections. In the s-channel configuration, the parameter $K$-factor takes a value of $\frac{10.32}{6.902}=1.5$.

\begin{figure}[!htb] 
\centering 
\includegraphics[width=0.495\textwidth]{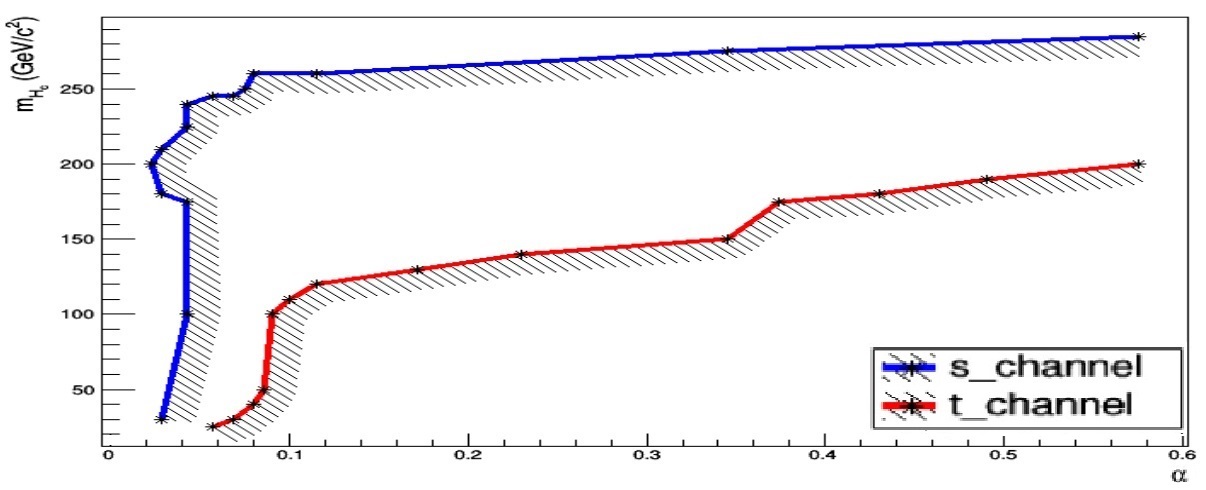} 
\includegraphics[width=0.495\textwidth]{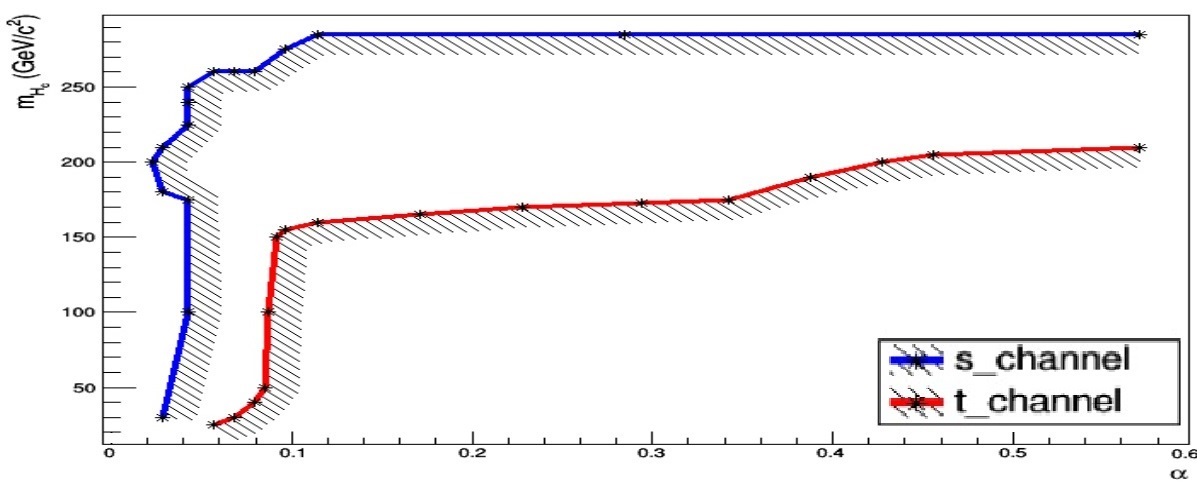} 
\caption{Allowed and excluded regions based on compatibility with experimental measurements (left) and precise agreement with SM predictions (right) for the $s$- and $t$-channels.} 
\label{fig:SigmaConstraints} 
\end{figure}

It is evident that the measured value has a relatively large uncertainty, which is expected to be improved in future, when more data is collected. To anticipate the impact of a more precise measurement, a second allowed range is defined. In this case, it is assumed that the measured cross-section value matches the SM prediction, with a measurement precision of 10\% of the predicted value, specifically:
\begin{equation} 
0.9~\sigma_{\text{SM($t+\bar{t}$)}} < \sigma_{\text{measured($t+\bar{t}$)}} < 1.1~ \sigma_{\text{SM($t+\bar{t}$)}}.
\label {eq:SigmaConstraints} 
\end{equation} 
For the \( t \)-channels, the second allowed range is defined similarly to the \( s \)-channels. For the first allowed range, the experimental measurement of the single top quark production cross-section in the \( t \)-channels, obtained by the CMS collaboration in 2017 with a value of 
$\sigma_{\text{t-channel}} = 238 \pm 31.78$ pb \cite{CMS:2016lel}, is used. The most precise SM prediction for this value is 217 pb \cite{CMS:2016lel}, compared to the software prediction of 179.91 pb, defining the $K$-factor = $\frac{217}{179.91} = 1.21$
In Fig.\ref{fig:SigmaConstraints}
the boundaries between the allowed and excluded regions of the introduced model are depicted. The results for the \( t \) and \( s \)-channels, are plotted in the $\alpha$ and \(m_{H_c}\) plane. An allowed point is defined as a point where the cross-section value falls within 1$\sigma$ of the measured/predicted central value. 

As expected and is illustrated in the figures, large coupling coefficients and low masses are excluded within the assumptions of our effective model. 
Within the examined regions, the allowed range for the $t$-channel is broader in both approaches, indicating that the constraining power of the \( s\)-channel is stronger than that of the \( t \)-channel. In other words, the excluded regions of the \( t \)-channel are a subset of the excluded regions of the \( s \)-channel. Increasing the collected data, can improve the results for low $\alpha$ and high $m_{H_c}$, without important changes in other parts of the phase space (Fig.\ref{fig:SigmaConstraints} (right)).
 
\subsection{Spin Correlation}
\begin{figure}[!htb] 
\centering 
\includegraphics[width=0.5\textwidth]{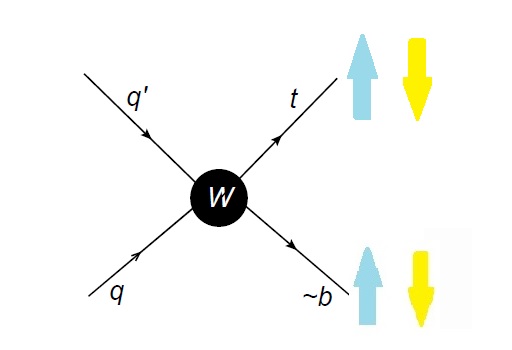} 
\caption{Different spin orientation of the $t$ quark and the \(\bar{b}\) quark in the rest frame of the $W$ boson. The yellow and blue arrows indicate the spin directions of the particles.} 
\label{fig:restw}
\end{figure}

\begin{figure}[htbp] 
\centering 
\includegraphics[width=0.5\textwidth]{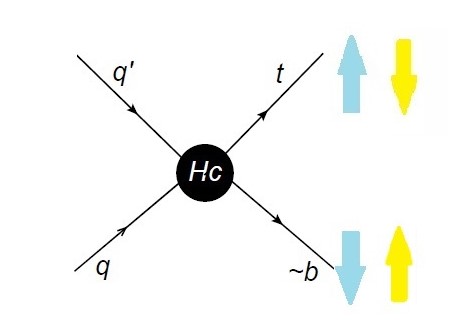} 
\caption{Different spin orientation of the $t$ quark and the \(\bar{b}\) quark in the rest frame of the $H_c$ boson.}
\label{fig:resth} 
\end{figure}

In the second approach, the difference in the spin of this charged scalar compared to the spin of the $W$ boson is considered.
The effect of the presence of a charged scalar as a mediator in the single top quark production process via the $s$-channel is investigated.
In SM, the mediator particle in the single top quark production process via the \( s \)-channel is the \( W \) vector boson. Consequently, for the spins of the \( W \) boson and the outgoing particles, the following relation exists. The $z$-direction is the momentum direction of the $W$ boson.
\begin{eqnarray}
W(S_z) = 1 \Rightarrow spin_t = \uparrow&,& spin_{\bar{b}} = \uparrow \\  \nonumber
W(S_z) = -1 \Rightarrow  spin_t = \downarrow&,& spin_{\bar{b}} = \downarrow \\ W(S_z) = 0 \Rightarrow  spin_t = \downarrow, spin_{\bar{b}} = \uparrow &\text{or}&  spin_t = \uparrow, spin_{\bar{b}} = \downarrow
\label {eq:spinw} 
\end{eqnarray}
In the above relation, the direction of the arrow indicates the spin orientation of the particle. The spins are shown in Fig\ref{fig:restw}, schematically.
According to the figure and relation, the spin direction of the top quark and the \(\bar{b}\) quark will be, usually, identical to that of the \( W \)  boson.
If the charged scalar particle, \( H_c \), is considered as the mediator in the single top quark production process, then according to Fig. \ref{fig:resth}, 
the following relation holds for the spins of the outgoing particles and the $H_c$ boson: 
\begin{equation} 
H_c(S_z = 0) \Rightarrow spin_t = \downarrow, spin_{b} = \uparrow \quad \text{or} \quad spin_t = \uparrow, spin_{b} = \downarrow 
\label {eq:spinh} 
\end{equation} 
In the above relation, the direction of the arrow indicates the spin orientation of the particle. 
The weak interactions, e.g. $W$ boson interactions, couple only to left-handed particles and right-handed antiparticles, making chirality a key feature in electroweak top quark production. But the new charged scalar, can couple to all helicities, (Look at Eq.\ref{eq:hcL}), so different spin orientations are expected for the outgoing  particles from scalar or vector mediators.

To quantify the effect of adding a new scalar mediator, the quantities are defined in the rest frame of the mediator. The cosine of the angle between the momentums of the outgoing top quark and the incoming \( q \) quark is examined. In the generated sample, the incoming \( q \) quark can only be a \( u \) or  \( c \) quark, so meaning that the sample contains only top quarks, without any antitop quark. Since the \( q \) quark, usually, originates from the valence quarks in the proton, it is therefore more energetic than the antiquark, which always arises from the sea quarks. Consequently, given the boost of the top and \( \bar{b} \)  quark system, it is expected that the direction of motion of the  \( q \) quark can be determined, even in real data.

\begin{figure}[htbp] 
\centering 
\includegraphics[width=0.48\textwidth]{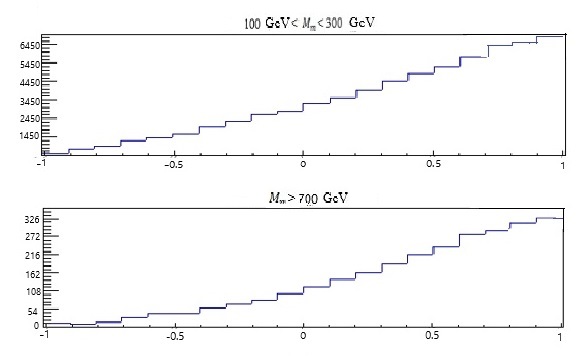} 
\includegraphics[width=0.48\textwidth]{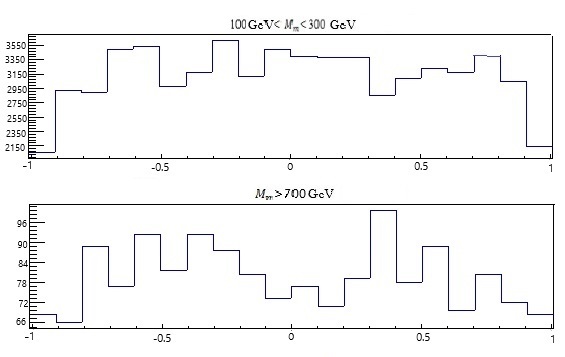} 
\caption{Cosine of the angle between the top quark and the incoming $q$ quark for SM (left) and the considered model with parameters \( (\alpha = 0.58 \mathrm{(GeV)}^{-1}, m_{H_c} = 100 \text{GeV}) \), without $W$ boson contribution (right). The angle is measured in the mediator rest frame.} 
\label{fig:cosine} 
\end{figure}

When the invariant mass of the outgoing top quark and \(\bar{b}\) system, 
denoted as \(M_{m}\) (mediator mass), is between 100 and 300 GeV, $100<M_m<300$~GeV, the sample is dominated by SM events containing $W$ boson as the mediator. It defines the control region. The signal region is defined when the mediator mass is greater than 700 GeV, $M_m > 700$ GeV. The signal region is expected to have more $H_c$ contribution. 
Figure \ref{fig:cosine}
illustrates the cosine of the angle between the directions of motion of the $q$ quark and top quark, in the mediator rest frame. As it is expected for SM, cosine tends to one, but for pure $H_c$ contribution, distribution of cosine tends to be flat. If the new model exists and $H_c$ contributes to SM, sum of the differently weighted histograms is seen in the experiment.

\begin{figure}[!htb] 
\centering 
\includegraphics[width=0.48\textwidth]{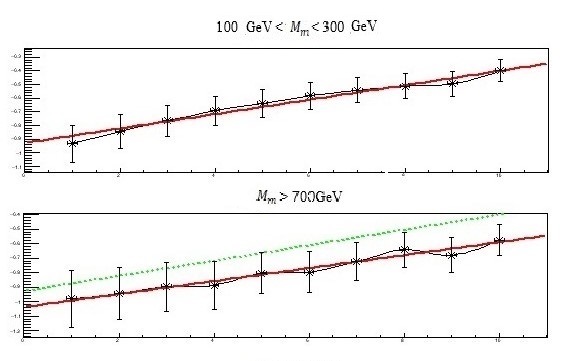} 
\includegraphics[width=0.48\textwidth]{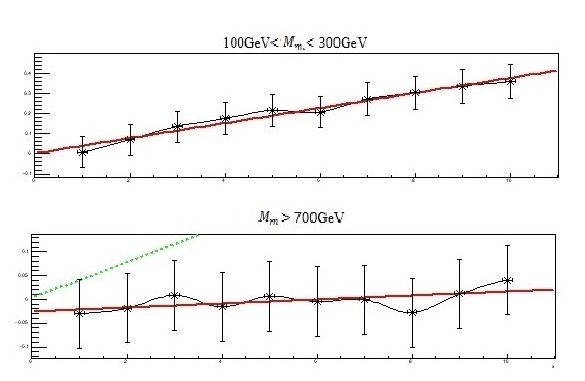} 
\caption{The graphs made for SM (left) and SM+$H_c$ (right) by the recipe explained in the text. The red lines represent the fitted lines for the control region and signal region. The green lines indicate the fitted line from the control region, refitted to the signal region, when the slope and intercept are fixed. The parameters of the new model are \(m_{H_{c}} = 300 \text{ GeV}\) and \(\alpha = 0.57\).} 
\label{fig:fit}
\end{figure}

To highlight these differences, the horizontal axis of each histogram is divided into twenty equal sub-intervals (bins). Then, using the following formula,
\begin {equation} 
y(i) = \frac{\text{bin } (i) - \text{bin} (i+10)}{\text{bin } (i) + \text{bin} (i+10)},
\label {eq:yi} 
\end {equation}
the pairs of \( (i,y(i)) \) are used to make the corresponding graphs of each histogram.
In this relation, the variable \( i \) varies sequentially from 1 to 10. To quantify the differences between the graphs in SM and the considered model, the graphs for control and signal regions are fitted using the first-degree polynomial (straight line). Then, the fitted line of the graph of the control region, with the fixed slope and intercept, is refitted to the graph of the signal region. Drawing and fitting the plots is done using the utilities of the ROOT data analysis framework \cite{Brun:1997pa}. 
As shown in Fig. \ref{fig:fit},
the line obtained from fitting the control region is consistent with signal region in SM. When the new model is added, this consistency is significantly lower in the signal region. The error bars shown in the figure, are the quadratic sum of the statistical and 10\% systematical uncertainties. The alternative values of 5 and 20\% are also examined. They change the boundaries, but the main message of the analysis remains without changes.

\begin{figure}[!htb]
\centering
\includegraphics[width=0.48\textwidth]{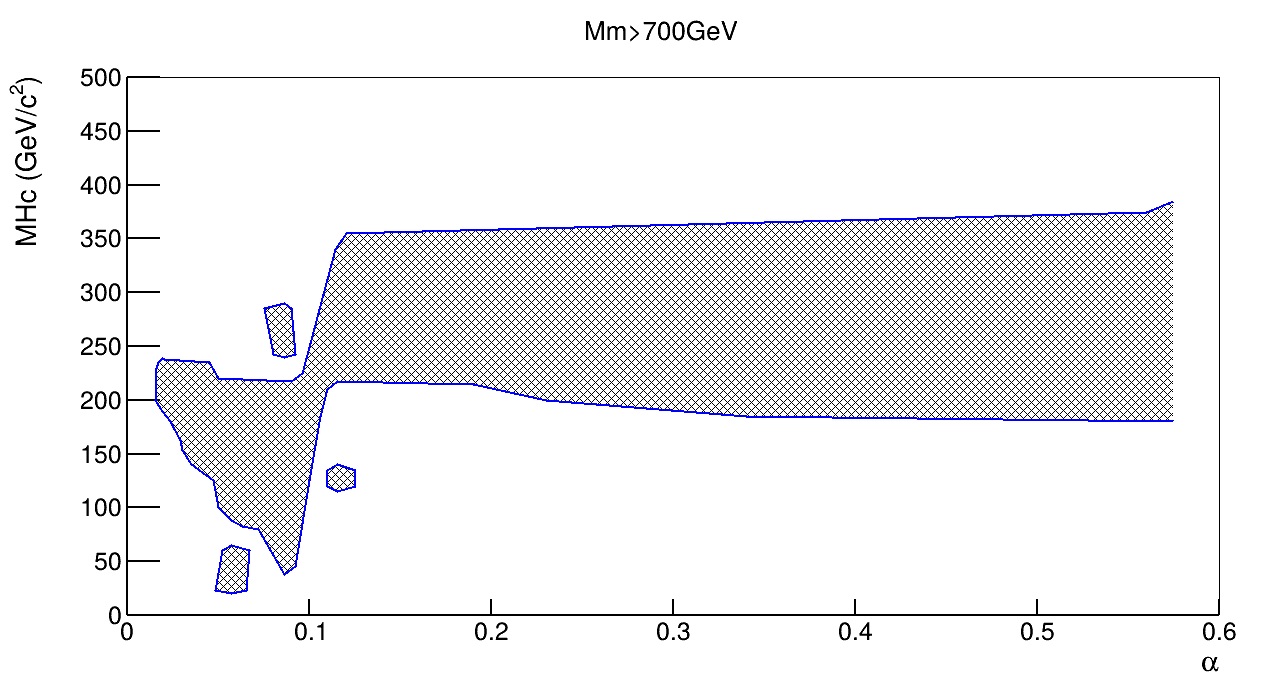}
%
\includegraphics[width=0.48\textwidth]{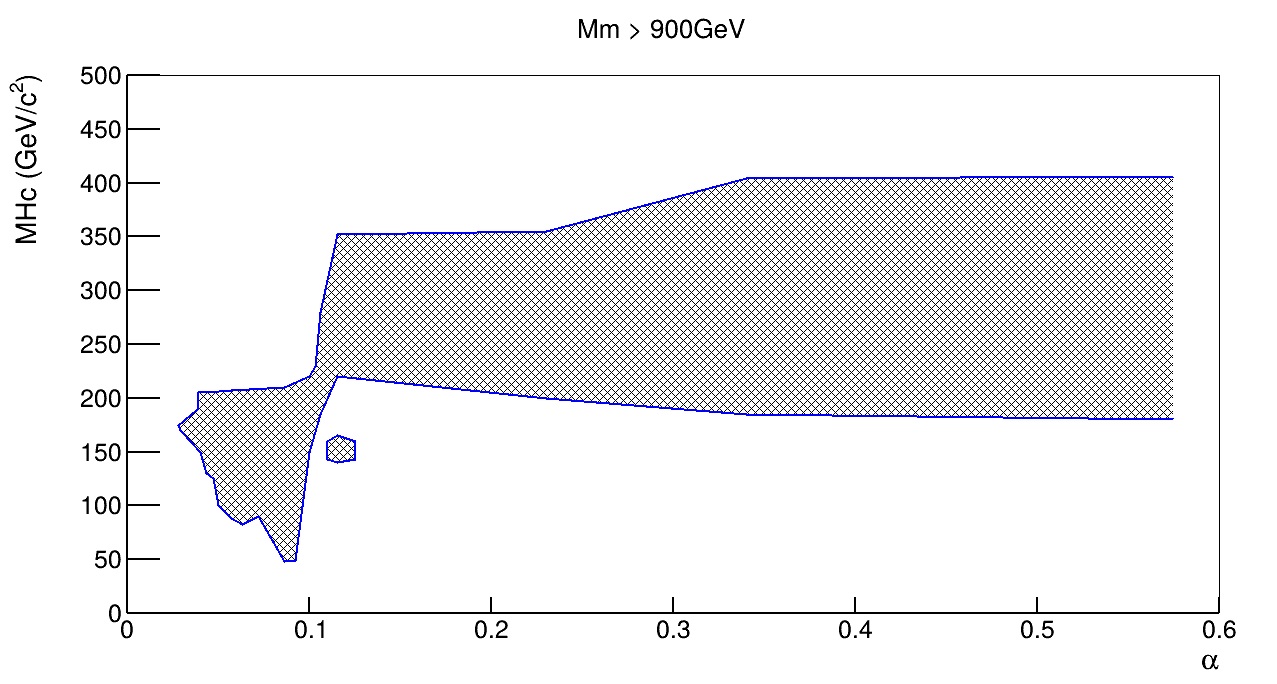}
\caption{The excluded region in the new model parameter space when the signal region is defined as \( M_m > 700 \) (left) and \( M_m > 900 \) (rifght).}
\label{fig:chi2prob}
\end{figure}

To scan the model's parameter space, \(\alpha \) is considered within the range \( [ 0 , 0.6 ]\) in 18 equal steps. The mass of the charged scalar is taken from 50 GeV to 450 GeV in steps of 50 GeV. 
Statistically, the chi-squared  ($\chi^2$) value in SM is much smaller than the $\chi^2$ value in the presence of the new model.  The probability of $\chi^2$ with 10 degrees-of-freedom serves as a criterion to constrain the new model.
For various points in the parameter space of the new model, the 
points with a $\chi^2$ probability of less than 0.05 are considered as excluded regions at a 95\% Confidence Level. This statistical analysis follows the method explained in Ref \cite{feldman1998unified}.
Based on this criterion, points with \( \chi^2 > 18.30 \) are identified as the excluded regions. Obviously, points with \( \chi^2 < 18.30 \) are classified as the allowed regions. The results for different points of the parameter space are shown in Fig. \ref{fig:chi2prob} (left).
It can be seen that for low values of $\alpha$ ($\alpha$ < 0.1), low masses ($50 \lesssim M_{H_c} \lesssim 200$ GeV) are excluded. For $\alpha$ > 0.1, medium masses ($200 \lesssim M_{H_c} \lesssim 350$ GeV) are excluded. To check the sensitivity to lower bound of the signal region, the results for \( M_m > 900 \) are shown in Fig. \ref{fig:chi2prob} (right). The overall picture, remains untouched. But, the exclusion gets weaker in low $\alpha$ and stronger in high $\alpha$. For high values of $\alpha$, masses close to 400 GeV are also excluded. 

\begin{figure}[!htb]
	\centering
	\includegraphics[width=0.48\textwidth]{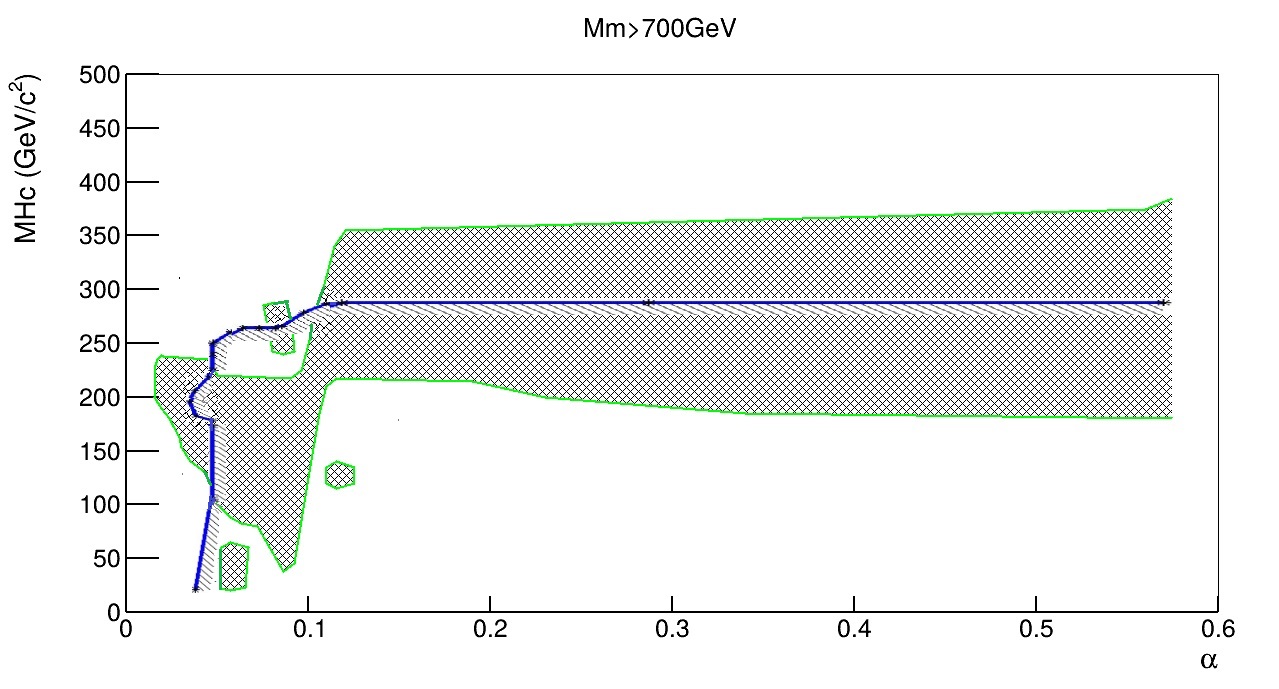}
	\includegraphics[width=0.48\textwidth]{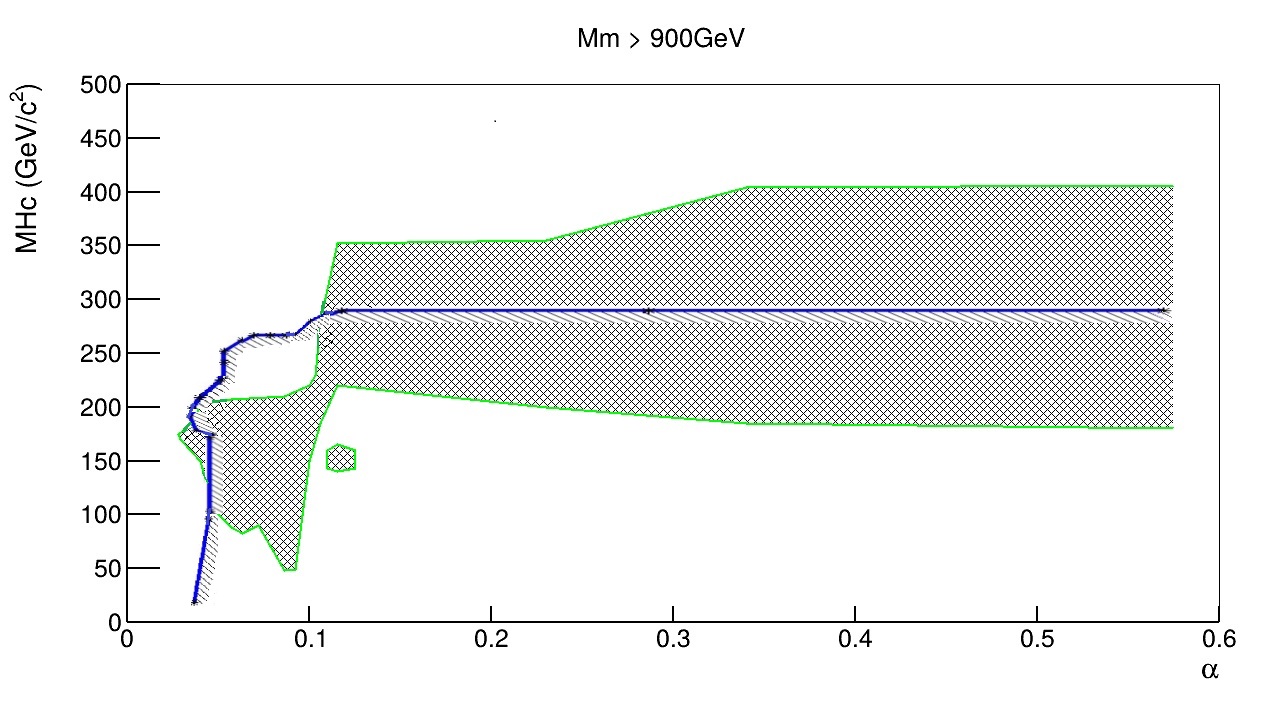}
	\caption{A comparison of the results obtained from two different approaches.}
	\label{fig:compare}
\end{figure}

To compare the two approaches, the best result of the first approach, the blue line in Fig.\ref{fig:SigmaConstraints} (right), is set on top of the plots of Fig. \ref{fig:chi2prob}. The result is shown in Fig.\ref{fig:compare}.
As can be seen, the first approach can rule out all low masses, except for very low $\alpha$. But, the second approach works for higher masses. Although it can not touch the low masses for high $\alpha$.

\begin{figure}[!htb]
	\centering
	\includegraphics[width=0.48\textwidth]{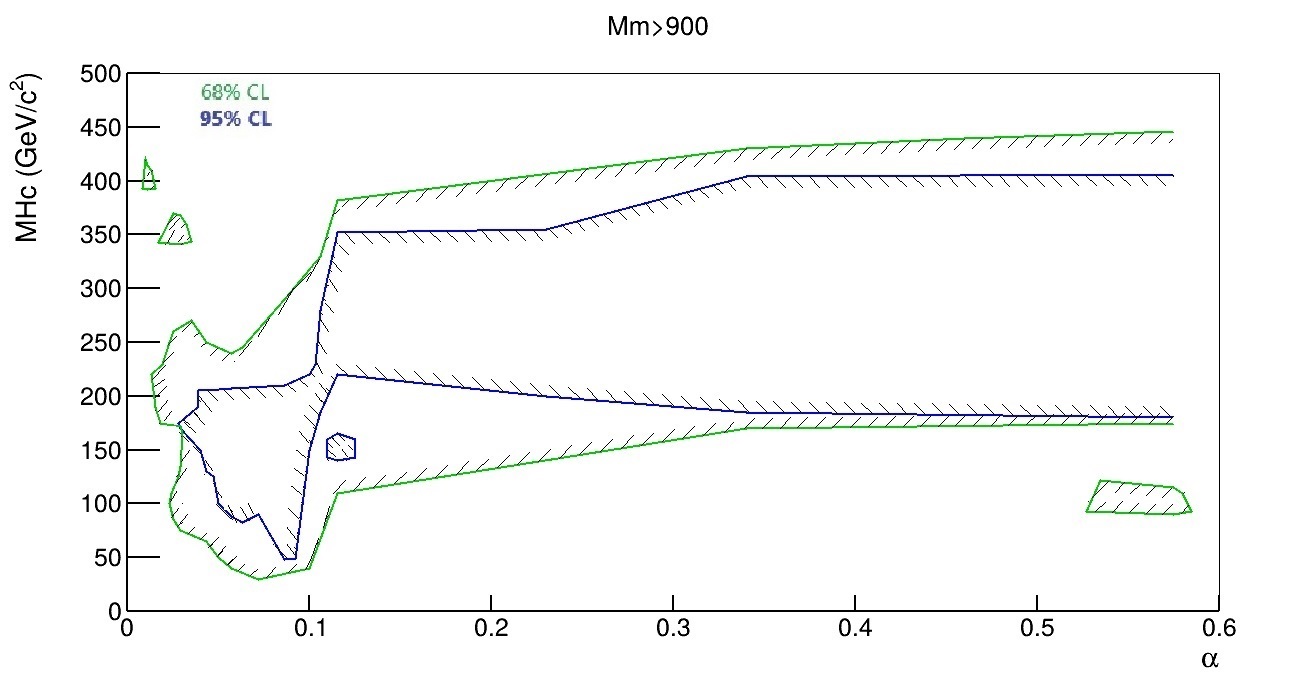}
	\caption{The exclusion areas for \( M_{m}\) > 900 evaluated at the 95\%  and 68\% confidence levels.}
	\label{fig:compare68and95}
\end{figure}

The Fig.\ref{fig:compare68and95} presents a comparison between the 95\% and 68\% 
confidence levels for $M_m$ values exceeding 900 GeV. As it is expected, the 95\% area is a subset of the 68\% area.

\subsection{Method Examination} 
To optimize the separation between the new model and SM, at each step of the algorithm used above, various alternative methods are also examined. Ultimately, the reported method is chosen based on the criterion of observing a larger excluded regions. Prior to final selection, investigations are conducted on the cosine of the angle between the directions of the outgoing \(t\) quark and the \(\bar{b}\) quark. 
Additionally, in Equation \ref{eq:yi}, instead of \(\text{bin}(i+10)\), the case of \(\text{bin}(21-i)\) is also considered. 
To define the control region, different definitions like \(300\) GeV \(< M_m < 500\) GeV, \(500\) GeV \(< M_m < 700\) GeV are also tried. As it was shown previously, the signal region can be defined as  $M_m > 900$ GeV, alternatively.
However, for all the mentioned variations, the results for the selected method are superior, so the alternatives are not reported here. 

\section{Conclusion}\label{sec.con}
In this paper, a new charged scalar particle, $H_c$, is added to the Standard Model. The effects of this new particle on single top quark production are investigated through two different approaches. In the first approach, the cross-section of single top quark production in the $s$- and $t$-channels is studied.
The results show that, within the simplified phenomenological framework adopted in this analysis, large parts of the parameter space, especially for low masses and large couplings, are disfavored by comparing the calculated cross-sections with the experimental measurements and SM predictions.
In the second approach, the spin correlation in the $s$-channel single top quark production is analyzed. Since the spin of the new scalar mediator is different from the $W$ boson in SM, different angular distributions of the final state particles are expected. By comparing the angular distributions in the control and signal regions, another set of constraints on the model parameters is obtained.
The combination of these two approaches shows that they are complementary. While the cross-section measurement is powerful to exclude low mass regions, the spin correlation analysis is able to probe higher mass regions effectively.
Therefore, within the scope of our effective description and under the theoretical and experimental assumptions discussed in Sec. II, using both methods together constrains a significant portion of the parameter space of the new model. We note that these constraints are subject to the simplified phenomenological assumptions of our analysis, and a more comprehensive study incorporating all theoretical and experimental bounds would be required to establish fully viable exclusion regions. 
Future more precise measurements at the LHC can improve these constraints and help to find any possible signature of such a new charged scalar particle.

\section{Acknowledgments}
S. P. would like to thank CERN TH-Devision for their support and kind hospitality during his stay at CERN.
%

\bibliography{top_2HDMBib}


\end{document}